\documentclass[aps,prb,showpacs,twocolumn]{revtex4}
\usepackage{graphics}
\usepackage{graphicx}
\usepackage{dcolumn} 
\usepackage{bm}
\usepackage{epsfig}
\newcommand{\hem}{Fe$_2$O$_3$}
\newcommand{\alo}{Al$_2$O$_3$}
\newcommand{\ahem}{$\alpha$-Fe$_2$O$_3$}
\newcommand{\ilm}{FeTiO$_3$}
\newcommand{\cro}{Cr$_2$O$_3$}

\newcommand{\fetw}{Fe$^{2+}$}
\newcommand{\feth}{Fe$^{3+}$}
\newcommand{\tith}{Ti$^{3+}$}
\newcommand{\tif}{Ti$^{4+}$}
\newcommand{\otw}{O$^{2-}$}
\newcommand{\eg}{{\sl e.g.}}
\newcommand{\etal}{{\sl et al.}}

\newcommand{\alhem}{a$_{\rm Fe_2O_3}$}
\newcommand{\alalo}{a$_{\rm Al_2O_3}$}
\newcommand{\alilm}{a$_{\rm FeTiO_3}$}

\newcommand{\lmh}{{\sl lamellar magnetism hypothesis}}

\begin{document}
\title{Effect of strain on the stability and electronic properties of ferrimagnetic Fe$_{2-x}$Ti$_x$O$_3$ heterostructures from correlated band theory}
\author{Hasan Sadat Nabi and Rossitza Pentcheva}
\email{pentcheva@lrz.uni-muenchen.de}
\affiliation{Department of Earth and Environmental Sciences,
University of Munich, Theresienstr. 41, 80333 Munich, Germany}
\date{\today}
\begin{abstract}
Based on density functional theory (DFT) calculations including an on-site Hubbard $U$ term we investigate the effect of substrate-induced strain on the properties of ferrimagnetic Fe$_2$O$_3$-FeTiO$_3$ solid solutions and heterostructures. While the charge compensation mechanism through formation of a mixed \fetw, \feth-contact layer is unaffected, strain can be used to tune the electronic properties of the system, \eg\ by changing the position of impurity levels in the band gap. Straining hematite/ilmenite films at the lateral parameters of Al$_{2}$O$_{3}$(0001), commonly used as a substrate, is found to be energetically unfavorable as compared to films on Fe$_{2}$O$_{3}$(0001) or FeTiO$_{3}$(0001)-substrates.
\end{abstract}
\maketitle

\section{Introduction}
In the fabrication of ferromagnetic semiconductors for spintronics applications a lot of research focuses on the homogeneous doping of traditional or oxide semiconductors with $3d$ ions~\cite{Coey05,MacDonald05,ohno1998,Matsumoto01}. However, the coupling between magnetic impurities and charge carriers is often too weak, leading to Curie temperatures ($T_C$) way below room temperature (RT). On the other hand, materials like Fe$_{2-x}$Ti$_x$O$_3$ exhibit intrinsic semiconducting and ferrimagnetic properties, although the end members \ahem\ and \ilm\ are antiferromagnetic insulators with $T_{\rm N}=948$ and 56~K, respectively. Besides applications in spintronics, this material is also discussed in paleomagnetism as a possible cause of anomalies in the Earth's magnetic field, as well as for electronics devices (\eg\ varistors) because it is a wide band gap semiconductor that can be either $n$- or $p$-type depending on the doping concentration~\cite{Zhou02}. A Curie temperature  above RT and a reduction of resistivity was observed in synthetic solid solutions with Ti concentrations up to 70\%~\cite{ishikawa57,ishikawa58}. Moreover, $T_C$ was found to increase upon annealing both in these samples and in thin epitaxial films~\cite{hojo06}.  This behavior can be attributed to cation ordering phenomena related to a miscibility gap in the rather complex phase diagram of the system~\cite{Robinson04}.

\begin{figure}[b]
\begin{center}
  \rotatebox{0.}{ \includegraphics[scale=0.39]{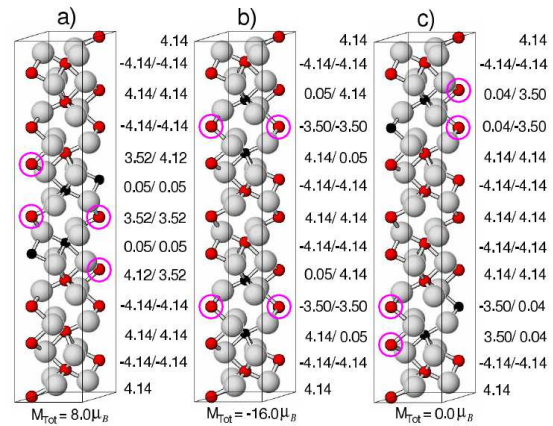}}
\end{center}
\caption{\label{fig:struct}  (Color online) Crystal structure of the 60-atom unit cell of Fe$_{2-x}$Ti$_x$O$_3$ for $x=0.33$ with a layered (a) and more homogeneous arrangement of the Ti-cations with Ti in the same (b) and  different (c) spin-sublattices. Oxygen, Fe and Ti are shown with light grey, red and black spheres respectively. Pink circles mark the \fetw-positions, while the rest of the iron are \feth. The local magnetic moments at the cation sites and the total magnetization of the system in $\mu_{\rm B}$ are given in the right side and bottom of each configuration, respectively.}
\end{figure}

The origin of ferrimagnetic behavior remained unclear until recently. Both materials have a corundum(-related) structure (see Fig.~\ref{fig:struct}) with a stacking of 2\feth/3\otw\ in hematite (space group $R\bar{3}c$) and 2\fetw/3\otw/2\tif/3\otw\ in ilmenite ($R\bar{3}$) along the [0001]-direction. Thus at an interface or in a solid solution (SS) charge is not compensated, if all ions preserved their bulk valence states. DFT calculations considering correlation effects within LDA+U~\cite{anisimov93} showed that the charge mismatch is accommodated by a mixed \feth, \fetw\ contact layer at the interface~\cite{Pentcheva08}, providing first theoretical evidence for the \lmh~\cite{nature04}. The \fetw-ions at the interface give rise to uncompensated moments  and also to impurity states in the band gap.

The incorporation of Ti in hematite~\cite{attfield91} ($a=5.04$ \AA, $c=13.75$ \AA) introduces a substantial strain: the volume of the end member ilmenite~\cite{harrison00} ($a=5.18$ \AA, $c=14.27$ \AA) is 9.7\% larger than the one of hematite. Indeed, lens-shaped dark contrasts around nanoscale hematite lamellae in an ilmenite host, imaged by transmission electron microscopy, indicate significant strain fields~\cite{nature04}.

Epitaxial Fe$_{2-x}$Ti$_x$O$_3$ films~\cite{Zhou02,Fujii04,hojo06,Kuroda07,popova08,ndilimabaka08} are typically grown on an Al$_{2}$O$_{3}$(0001)-substrate ($a=4.76$~\AA, $c=12.99$~\AA) which introduces a substantial compressive strain of 5.8\% and 8.8\% compared to \hem\ and \ilm\ and only rarely, a \cro-buffer layer is used~\cite{Chambers06} to reduce the lattice mismatch.

Epitaxial strain can have a strong impact on the film properties, \eg\ by tuning the magnetic interactions in magnetoelastic composites~\cite{Eerenstein06}, enhancing ferroelectricity ~\cite{choi2004, zhang2007} or even inducing orbital reconstructions~\cite{izumi01}. The goal of the present study is to explore the effect of strain on the properties of Fe$_{2-x}$Ti$_x$O$_3$. In particular we address its influence on $(i)$ the energetic stability and compensation mechanism as well as on $(ii)$ the electronic, magnetic and structural properties of the system. DFT calculations are performed on SS and layered configurations with $x=0.17, 0.33, 0.50$ and 0.66, strained laterally at the lattice parameters of Al$_{2}$O$_{3}$, Fe$_{2}$O$_{3}$, and FeTiO$_{3}$.

\section{Calculational details}
We use the all-electron full-potential linear augmented plane wave (FP-LAPW) method as implemented in the WIEN2K code~\cite{wien} and the generalized gradient approximation (GGA)~\cite{pbe96}. Within LDA+U~\cite{anisimov93} $U=8.0$~eV and $J=1.0$~eV is applied to the Fe and Ti $3d$ states. These values were found to reproduce correctly the ground state of \ilm~\cite{Pentcheva08}. The systems are simulated in a hexagonal unit cell with 60 atoms (Fig.~\ref{fig:struct}). Besides the layered configurations (cf. Fig.~\ref{fig:struct}a) more homogeneous distributions are generated by substituting $50\%$ of Fe in a bilayer by Ti, as shown \eg\ in Fig.~\ref{fig:struct}b-c. For further details on the calculation see (Ref.~\cite{Pentcheva08}).

\begin{figure}[t!]
\begin{center}
\rotatebox{0.}{
\includegraphics[width=2.8in]{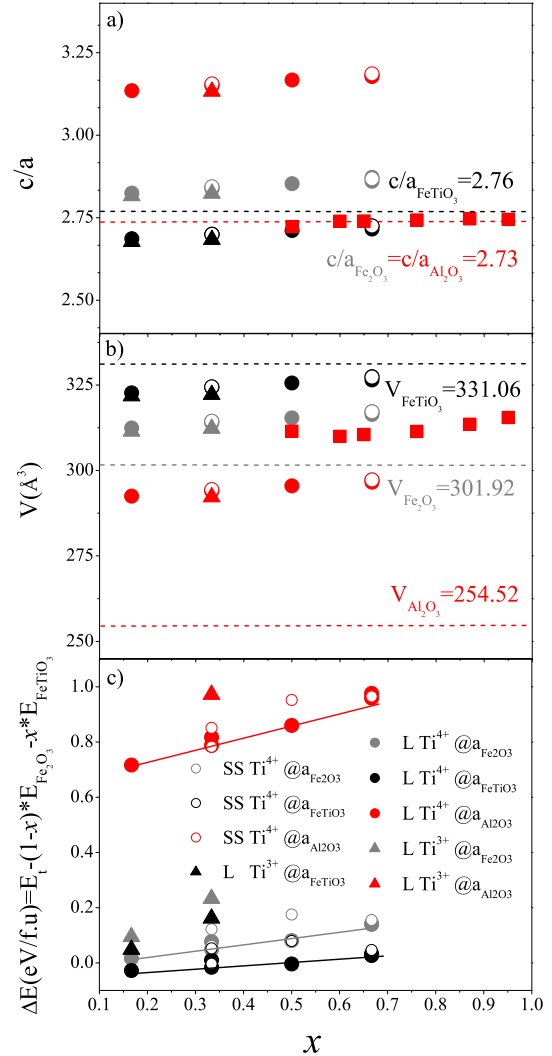}}
\end{center}
\caption{\label{c-vers-x} (Color online)
(a) $c/a$-ratio (b) volume and (c) formation energy (eV/f.u) versus ilmenite concentration $x_{\rm Ilm}$ for Fe$_{2-x}$Ti$_x$O$_3$ strained at the Al$_{2}$O$_{3}$ (red/dark grey), Fe$_{2}$O$_{3}$ (grey) and FeTiO$_{3}$ (black) lateral lattice constants. Circles (triangles) denote compensation involving \tif (\tith). Open/filled symbols refer to solid solutions (SS)/ layered configurations (L).  Horizontal lines mark the bulk $c/a$ ratio and volume of the end members and \alo. Red (dark grey) squares indicate experimental data from
Takada~\etal~\cite{takada08}.}
\end{figure}
\section{Results and Discussion}
The optimized $c/a$-ratio and volume (Fig.~\ref{c-vers-x}a-b) show a linear increase with $x_{\rm Ilm}$ in accordance with Vegard's law, similar to what was observed experimentally in synthetic hematite-ilmenite solid solutions~\cite{ishikawa57}. Furthermore, for a given concentration both $c/a$ and $V$ are largely independent of the distribution of Ti-impurities.
The $c/a$-ratio of bulk \ilm\ (2.76) is slightly larger  than the one for \ahem\ and \alo\ (2.73). Due to the small tensile/compressive strain when using \alilm/\alhem~the $c/a$-ratio of Fe$_{2-x}$Ti$_x$O$_3$ is slightly reduced (-1.1 to -2.8 \%)/increased  (3.1-5.2 \%), respectively. In contrast, due to the high compressive strain on an \alo-substrate, $c/a$ increases strongly by 14.7-16.6 \% which corresponds to $c_{rel}=14.89-15.15$~\AA. Nevertheless, the volume does not completely relax: The volume of the system strained at the \alo-lateral lattice constant is 6.8 \% (10.2 \%) smaller than when strained at \alhem\ (\alilm). The volumes of Fe$_{2-x}$Ti$_x$O$_3$ strained at \alilm\ and \alhem\ lie between the ones for the end members \hem\ and \ilm.

X-ray diffraction data for Fe$_{2-x}$Ti$_x$O$_3$ films on \alo(0001)~\cite{takada08,popova08-1} indicate significant lateral strain relaxation: already in a 10~nm thick film $a$ relaxes to the bulk value of \ilm\ with only a small change in $c/a$ (see Fig.~\ref{c-vers-x}a). The $c/a$ values and volumes obtained by Takada~\etal~\cite{takada08} are in good agreement with the DFT values of the systems strained at \alilm.

Next we turn to the influence of strain on the energetic stability. The formation energy with respect to the end members as a function of $x_{\rm Ti}$ is shown in Fig.~\ref{c-vers-x}c for the three different substrate lattice constants. For each Ti-concentration we have considered several different cation arrangements, \eg\ for $x=0.33$ these include an ordered arrangement with an Fe layer sandwiched between two Ti layers (Fig.~\ref{fig:struct}a) or solid solutions with Ti ions either in the same (Fig.~\ref{fig:struct}b) or different spin-sublattices (Fig.~\ref{fig:struct}c). We find that compensation through \tif\ and disproportionation in \fetw, \feth\ is more favorable over mechanisms involving \tith. Furthermore, the formation energy increases linearly with $x_{\rm Ilm}$. These features are independent of the substrate lattice parameters. Systems strained laterally at \alilm\ are more stable than the ones on \alhem. In contrast, the formation energy of films strained at \alalo\ increases by 0.7~eV as compared to films on \alhem. This implies that the strong compressive strain is energetically unfavorable and gives a possible explanation why a lateral strain relaxation occurs in Fe$_{2-x}$Ti$_x$O$_3$ films~\cite{takada08,popova08-1}.
While for systems strained on hematite and ilmenite substrates layered arrangements (full symbols) are more favorable than homogeneous distributions (open symbols), the trend is reversed for $x=0.33$ and  $x=0.66$ on an  \alo(0001)-substrate.

\begin{figure}[t]
\begin{center}
\rotatebox{0.}{
\includegraphics[width=3.in]{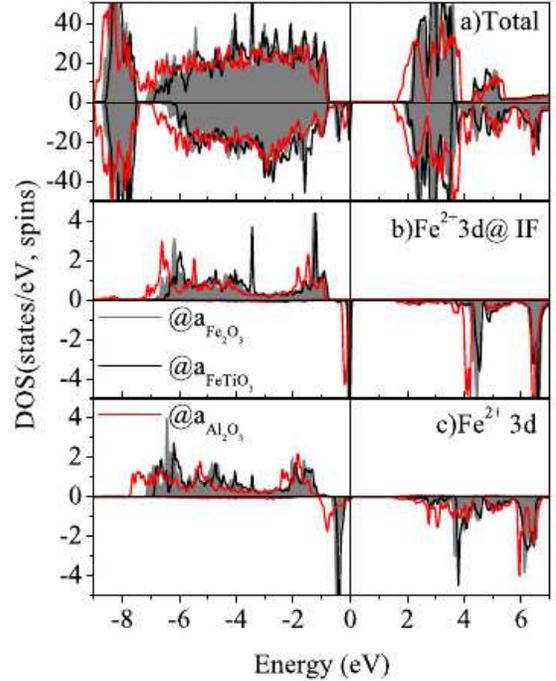}}
\end{center}
\caption{\label{dos} (Color online) Density of states  of Fe$_{1.67}$Ti$_{0.33}$O$_3$ containing two Ti-layers in a hematite host (shown in Fig.~\ref{fig:struct}a): a) total; b) and c) projected of the $3d$ states of \fetw\ at the interface and between the two Ti layers. The DOS of the system strained at the lateral lattice parameters of \hem, \alo, and \ilm\ is shown with a grey shaded area, red (dark grey), and black line, respectively. }

\end{figure}

Concerning the electronic properties of the hemo-ilmenite system, we have plotted in Fig.~\ref{dos} the density of states of a Ti-double layer in a hematite host (Fig.~\ref{fig:struct}a), but similar behavior is observed for all studied systems.  Upon \tif\ substitution, an iron-ion from the neighboring layer turns \fetw, as  observed also for isolated impurities by Velev~\etal~\cite{Velev05}. The \fetw O$_6$ and the TiO$_6$-octahedron are corner- (and not face-)sharing. The so formed \fetw-ions in the contact layer have an impurity state of $a_{1g}$ symmetry ($d_{z^2}$) that is pinned at the Fermi level for systems strained at \alhem\ and \alilm. Such a mid-gap state was recently reported from x-ray valence band photoemission~\cite{droubay07} and optical measurements~\cite{ndilimabaka08}, although it was related to the low oxygen pressure during deposition. The main feature related to strain is the change in band width: While for tensile strain at \alilm\ the bands are narrowed, for compressive strain at \alalo\ they are strongly broadened. This results in a reduction of the band gap (between the impurity state defining the Fermi level and the bottom of the conduction band)  from 1.90~eV for \alilm\ and 1.79~eV for \alhem\ to 1.43~eV for \alalo. The corresponding values for $x=66\%$ show the same trend but are smaller: 1.64~eV for \alilm\ and 1.46~eV for \alhem\ to 0.78~eV for \alalo.

The local magnetic moments and total magnetization for the three systems with $x=0.33$ is displayed in Fig.~\ref{fig:struct}. Strain has only a small impact on the magnetic moments of \fetw\ ($\sim 3.5~\mu_{\rm B}$) and \feth\ ($\sim 4.1~\mu_{\rm B}$) respectively which are reduced by less than 0.05 $\mu_{\rm B}$ at \alalo.
The \fetw-layer sandwiched between two Ti-layers in Fig.~\ref{fig:struct}a is only weakly coupled to the next Fe-layer (parallel and antiparallel orientation of the magnetic moments is nearly degenerate as in the ilmenite end member). Therefore, at temperatures above the N{\'e}el temperature of ilmenite, such layers will not contribute to the total magnetization. In contrast, \fetw\ in the contact layer shows a strong antiferromagnetic coupling to the neighboring  Fe-layer of the hematite host. These defect interface moments are responsible for the ferrimagnetic behavior of the system ($M_{tot}=8.0~\mu_{\rm B}$). In solid solutions, Ti substitution in different spin-sublattices (\eg~in adjacent layers as shown in Fig.~\ref{fig:struct}c), resulting in a zero net magnetization, is less favorable compared to substitution in the same spin-sublattice (Fig.~\ref{fig:struct}b), which maximizes the total magnetization ($M_{tot}=-16.0~\mu_{\rm B}$). This trend promotes  ferrimagnetic behavior in the system.
\section{Conclusions}
Density functional theory calculations within GGA+$U$ show that the charge compensation in hematite-ilmenite heterostructures and solid solutions  takes place through a mixed \fetw, \feth~contact layer. This mechanism is robust with respect to substrate-induced strain. For \hem(0001) or \ilm(0001) substrates layered arrangements  are more stable than solid solutions. However, the compressive strain at \alalo\ is likely to cause a stronger competition and even reverse the trend for $x=0.33$ and $x=0.66$. The growth of epitaxial films on an \alo-substrate is connected with a high energy cost. Therefore, in order to release strain such films may  roughen or buckle in the first layers as recently reported by Popova~\etal~\cite{popova08-1}. In contrast, the growth on lattice matched substrates or even substrates that produce a small tensile strain like \ilm\ is energetically favored.  Our DFT results indicate that strain can have a strong impact on the structural and electronic properties in the hematite-ilmenite system: \eg\ by tuning the band width or the position of impurity levels in the band gap and thus changing the concentration of spin-polarized carriers.
\section{Acknowledgments}
Funding by the DFG (Pe883/4-1), ESF (EuroMinSci) and computational time at the Leibniz Rechenzentrum are gratefully acknowledged.


\begin{thebibliography}{10}
\bibitem{ohno1998}H. Ohno, Science {\bf 281}, 951 (1998).

\bibitem{Matsumoto01}Y. Matsumoto, M. Murakami, T. Shono, T. Hasegawa, T. Fukumura, M. Kawasaki, P. Ahmet,
T. Chikyow, S. Koshihara, and H. Koinuma, Science {\bf 291}, 854 (2001).

\bibitem{Coey05} J. M. D. Coey, M. Venkatesan, and C. B. Fitzgerald, Nature Mater. {\bf 4}, 173 (2005).

\bibitem{MacDonald05} A. H. MacDonald, P. Schiffer, and N. Samarth, Nature Mater. {\bf 4}, 195 (2005).

\bibitem{Zhou02}F. Zhou, S. Kotru, and R. K. Pandey, Thin Solid Films, {\bf 408}, 33 (2002).

\bibitem{ishikawa57} Y. Ishikawa and S. Akimoto, J. Phys. Soc. Jpn. {\bf 12}, 1083 (1957).

\bibitem{ishikawa58} Y. Ishikawa, J. Phys. Soc. Jpn. {\bf 13}, 37 (1958).

\bibitem{hojo06} H. Hojo, K. Fujita, K. Tanaka, and K. Hirao, Appl. Phys. Lett.  {\bf 89}, 142503 (2006).

\bibitem{Robinson04} P. Robinson, R. J. Harrison, S. A. McEnroe, and R. B. Hargraves, Amer. Min. {\bf 89},725 (2004).

\bibitem{anisimov93} V. I. Anisimov, I. V. Solovyev, M. A. Korotin,
M. T. Czy\.{z}yk, and G. A. Sawatzky, Phys. Rev. B {\bf 48}, 16929 (1993).

\bibitem{Pentcheva08}R. Pentcheva and H. Sadat Nabi, Phys. Rev. B, {\bf 77}, 172405 (2008).

\bibitem{nature04} P. Robinson, R. J. Harrison, S. A. McEnroe, and R. B. Hargraves, Nature {\bf 418}, 517 (2002).

\bibitem{attfield91}D. A. Perkins and J. P. Attfield,  J. Chem. Soc., Chem. Commun., 229 (1991).

\bibitem{harrison00} R. J. Harrison, S. A. T. Redefern, and R. I. Smith, Am. Miner. {\bf 85}, 194 (2000).

\bibitem{Fujii04}T. Fujii, M. Kayano, Y. Takada, M. Nakanishi, and J. Takada, J. Mag. Mag. Mat. {\bf 272-276}, 2010 (2004).

\bibitem{popova08}E. Popova, B. Warot-Fonrose, H. Ndilimabaka, M. Bibes, N. Keller, B. Berini,
K. Bouzehouane, and Y. Dumont, J. Appl. Phys. {\bf 103}, 093909 (2008).

\bibitem{ndilimabaka08} H. Ndilimabaka, Y. Dumont, E. Popova, P. Desfonds, F. Jomard, N. Keller, M. Basletic, K. Bouzehouane, M. Bibes, and M. Godlewski, J. Appl. Phys. {\bf 103}, 07D137 (2008).

\bibitem{Kuroda07} S. Kuroda, N. Nishizawa, K. Takita, M. Mitome, Y. Bando, K. Osuch, and Tomasz Dietl,
Nature Mater. {\bf 6}, 440 (2007).

\bibitem{Chambers06} S. A. Chambers, T. C. Droubay, C. M. Wang, K. M. Rosso, S. M. Heald, D. A. Schwartz,
 K. R. Kittilstved, and D. R. Gamelin, Materials Today {\bf 9}, 28 (2006).

\bibitem{Eerenstein06} W. Eerenstein, J. F. Scott, and N. D. Mathur, Nature {\bf 442}, 759 (2006).

\bibitem{choi2004} K. J. Choi, M. Biegalski, Y. L. Li, A. Sharan, J. Schubert, R. Uecker, P. Reiche, Y. B. Chen,
 X. Q. Pan, V. Gopalan, L.-Q. Chen, D. G. Schlom, and C. B. Eom, Science {\bf 306}, 1005 (2004).

\bibitem{zhang2007} J. X. Zhang, Y. L. Li, Y. Wang, Z. K. Liu, L. Q. Chen, Y. H. Chu, F. Zavaliche,
and R. Ramesh, J. Appl. Phys. {\bf 101}, 114105 (2007).

\bibitem{izumi01} M. Izumi, Y. Ogimoto, Y. Konishi, T. Manako, M. Kawasaki, and Y. Tokura, Mat. Sci. Eng. B{\bf 84}, 53 (2001).

\bibitem{wien} P. Blaha,  K. Schwarz, G. K. H. Madsen, D. Kvasnicka, and J. Luitz, WIEN2k, An
Augmented Plane Wave+Local Orbitals Program for Calculating Crystal Properties
(Techn. Universit\"at Wien, Austria), 2001, ISBN 3-9501031-1-2.

\bibitem{pbe96} J. P. Perdew, K. Burke, and M. Ernzerhof, Phys. Rev. Lett. {\bf 77}, 3865, (1996).


\bibitem{popova08-1}E. Popova, H. Ndilimabaka, B. Warot-Fonrose, M. Bibes, N. Keller, B. Berini,
F. Jomard, K. Bouzehouane, and Y. Dumont, Appl. Phys. A {\bf 93}, 669 (2008).

\bibitem{takada08}Y. Takada, M. Nakanishi, T. Fujii, J. Takada, and Y. Muraoka, J. Appl. Phys. {\bf 104}, 033713 (2008).

\bibitem{Velev05} J. Velev, A. Bandyopadhyay, W. H. Butler, and S. K. Sarker, Phys. Rev. B {\bf 71}, 205208 (2005).

\bibitem{droubay07} T. Droubay, K. M. Rosso, S. M. Heald, D. E. McCready, C. M. Wang, and S. A. Chambers, Phys. Rev. B {\bf 75}, 104412 (2007).


\end{thebibliography}
\end{document}